# Effect of the Heliospheric State on CME Evolution

**Fithanegest Kassa Dagnew[1,2,3], Nat Gopalswamy[2], Solomon Belay Tessema[1], Sachiko Akiyama[2,3], Seiji Yashiro[2,3]**

[1] Ethiopian Space Science and Technology Institute (ESSTI), Entoto Observatory and Research Center (EORC), Addis Ababa, Ethiopia
[2] NASA Goddard Space Flight Center, Greenbelt, MD, USA
[3] The Catholic University of America, Washington DC, USA

## Abstract

The culmination of solar cycle 24 by the end of 2019 has created the opportunity to compare the differing properties of coronal mass ejections (CMEs) between two whole solar cycles: Solar cycle 23 (SC 23) and Solar cycle 24 (SC 24). We report on the width evolution of limb CMEs in SC 23 and 24 in order to test the suggestion by Gopalswamy et al. (2015a) that CME flux ropes attain pressure balance at larger heliocentric distances in SC 24. We measure CME width as a function of heliocentric distance for a significantly large number of limb CMEs (~1000) and determine the distances where the CMEs reach constant width in each cycle. We introduced a new parameter: the transition height ($hc$) of a CME defined as the critical heliocentric distance beyond which the CME width stabilizes to a quasi-constant value. Cycle and phase-to-phase comparisons are based on this new parameter. We find that the average value of $hc$ in SC 24 is 62% higher than in SC 23. SC 24 CMEs attain their peak width at larger distances from the Sun as compared to SC 23 CMEs. The enhanced transition height in SC 24 is new observational ratification of the anomalous expansion. The anomalous expansion of SC 24 CMEs which is caused by the weak state of the heliosphere, accounts for the larger heliocentric distance where the pressure balance between CME flux rope and the ambient medium is attained.

Unified Astronomy Thesaurus concepts: Solar coronal mass ejections (310), Active sun (18), Solar cycle (1487), Solar flares (1496), Heliosphere (711), Solar x-ray flares (1816)

Supporting material: data behind figure

# 1. Introduction

Solar cycle (SC) 24, the weakest since the commencement of the space age, caused a considerably weak state of the heliosphere due to the reduced solar activity. Heliospheric parameters of great significance in the interaction of the heliosphere with interstellar medium have substantially declined. The dynamic pressure, proton temperature, solar wind density, interplanetary magnetic field strength, momentum and energy flux, and Alfven speed significantly diminished in the inner heliosphere in SC 24 as compared to SC 23 (McComas et al. 2013; Gopalswamy et al. 2015a). The total external pressure declines with the distance from the Sun (Gopalswamy et al. 2014; Gopalswamy et al. 2015b), which sustains the expansion of coronal mass ejection (CME) flux ropes as they propagate into the interplanetary space and results in higher CME width in SC 24. A comparison of the expansion speeds of limb CMEs between cycles 23 and 24 proved that for a given CME radial speed, the expansion speed in SC 24 is significantly higher (Dagnew et al. 2020a). The abundance of halo CMEs (HCMEs) normalized to the sun spot number (SSN) is significantly higher in SC 24 in spite of the considerable drop in the SSN (Gopalswamy et al. 2015c; Dagnew et al. 2020b). The increased CME width, higher CME expansion speed, and enhanced abundance of HCMEs in SC 24 are all observational evidences of the anomalous expansion of SC 24 CMEs. The anomalous expansion of CMEs must occur very close to the sun before CMEs acquire constant widths in the coronagraph field of view (Gopalswamy et al. 2014). CME expansion is interconnected with pressure balance between the total pressure inside the flux rope and that in ambient medium (Moore et al. 2007; Gopalswamy et al. 2015a). Observations from SC 23 have shown that CME mass and width steadily increase until ~5Rs and then stabilize (Vourlidas et al. 2002; Gopalswamy 2004; Yashiro et al. 2004; Gopalswamy et al. 2015a). This height is thought to be the pressure balance height.

Magnetic clouds (MCs) are the interplanetary (IP) manifesations of CMEs characterized by a smooth large scale rotation of the magnetic field direction, strong magnetic field and low proton temperature (for example, Burlaga et al. 1981). Flux rope is the fundamental magnetic structure of a CME and it exhibits as an MC in in-situ observations. The CME flux rope is believed to be either preexisting or formed during the eruption process and is observed as an MC in the IP medium (for example, Gosling 1990; Leamon et al. 2004; Qiu et al. 2007; Gopalswamy et al. 2015d). The IP CMEs (ICMEs) may not appear as MCs due to observational limitations (Riley and Richardson 2013). The Observation of flux rope structure depends on the viewing angle of the in situ space craft: the structure is observed if the spacecraft passes through the nose of the ICMEs and may not be detected if it passes through the flanks (for example, Marubashi 1997; Gopalswamy 2010). Ever since Goldstein (1983) showed that MCs can be represented as force free, cylindrical magnetic flux ropes, MC and flux rope have been used conversely.

Gopalswamy et al. (2015a) found that the size of the SC-24 MCs at 1 AU is significantly smaller, contrary to the wider CMEs near the Sun. They explained a possible reason for this difference: narrow CMEs tend to appear like normal CMEs due to anomalous expansion which resulted in a substantial part of the SC 24 flux ropes commence as smaller sized ones. This can be inferred from the significantly lower CME mass in SC 24 which is also due to anomalous expansion that makes narrow CMEs appear like normal. For example, a $60^0$ wide CME in SC 24 is equivalent to a narrow one in SC 23 (Gopalswamy et al. 2015d). Based on the observed sizes at 1 AU in SCs 23 and 24, they suggested that the enhanced CME expansion in SC 24 near the Sun caused a larger heliocentric distance for the pressure balance in that SC. They arrived at this result by determining the ratio of the pressure balance distances between SC 23 and 24 considering the proportionality between CME widths at the Sun and the corresponding flux rope sizes at 1 au. They estimated the height where the pressure balance for SC 24 is reached much larger than the typical ~5 Rs in cycle 23 noted above.

SC 24 has come to its completion by the end of 2019. This has created an opportunity to compare the differing properties of CMEs between the two whole solar cycles (SC 23 and SC 24). In this study, we compare the width evolution of limb CMEs in the two whole solar cycles in order to test the suggestion by Gopalswamy et al. (2015a) that CME flux ropes attain pressure balance at larger heliocentric distances in SC 24. This is an important issue to understand the flux rope expansion in the heliosphere. For example, Lugaz et al. (2020) concluded that the flux rope expansion is due to the decrease in solar wind dynamic pressure with distance, contradicting an earlier work that shows the expansion is determined by the MC-to-ambient total pressure difference (Gopalswamy et al. 2015a). Yermolaev et al. (2021) reported that the reduction in the solar wind dynamic pressure in SC 24 is only by 14% relative to SC 23, so it is difficult to see the change in dynamic pressure causing the difference in the 1-au expansion speeds in the two cycles. The Gopalswamy et al. (2015a) prediction that SC 24 CMEs attain pressure balance at larger heliocentric distances is based on the MC-to-ambient total pressure difference in the two cycles, so verifying the prediction will clarify the cause of flux rope expansion at 1 au. In addition to comparing the width evolution between the two cycles, we also compare it in the corresponding phases (rise, maximum, and declining) of the two cycles. This work which demonstrates the impact of the heliospheric state on CME width evolution is another revelation of the diverse interaction between the heliosphere and CMEs in the two cycles.

## 2. Observations

The Large Angle and Spectrometric Coronagraph (LASCO) [Brueckner et al. 1995] on board the Solar and Heliospheric Observatory (SOHO) mission has been providing the most extensive CME images through its two externally occulted wide-ranging field of view (FOV) telescopes C2 and C3. It has generated a comprehensive, uniform, and continuous (with the exception from the loss of a few months of observation in 1998) critical data set essential for understanding CMEs and the long term eruptive behavior of the Sun (Gopalswamy et al. 2009). The Coordinated Data Analysis Workshop catalog (CDAW) [Yashiro et al. 2004; Gopalswamy et al. 2009] compiles all CMEs manually identified from SOHO/LASCO images within the C2 and C3 FOV. The catalog comprises parameters essential for the scientific investigations of CME properties and their consequences.

CMEs and solar flares are revelations of the physical process by which the Sun's accumulated magnetic energy is released through the process of magnetic reconnection. CMEs propagate in to the interplanetary space as large scale eruptions of magnetized plasma. Flares appear as a sudden flash of light on a relatively small scale with flare sites confined to the low solar atmosphere. CMEs interact with the heliosphere and their properties are affected by the heliospheric state, whereas flares are not (Gopalswamy et al. 2020a; Gopalswamy et al. 2020b). The solar flares are classified in to logarithmic X-ray classes A, B, C, M, and X based on the peak flux in watts per square meter (W/m$^2$) with wave lengths 1-8 Angstroms (Å) measured by the X-ray spectrometers (XRS) on board the GOES space craft where the peak fluxes ranging from <10$^{-7}$ (A) to ≥10$^{-4}$ (X) are divided in to a linear scale [1-9] (https://www.ngdc.noaa.gov/stp/solar/solarflares.html. https://www.stce.be/educational/classification).

We selected CMEs with soft X-ray flare size ≥ C3.0 (defined as flares with soft X-ray intensity in the 1-8 Å GOES energy channel ≥ 3.0×10$^{-6}$ W/m$^2$ according to the NOAA's Space

Weather Prediction Center) and within $30^0$ from the limb. Our selection criteria for the soft X ray flare size exclude the uncertainty in CME identification for weak flares. The choice of CMEs close to the solar limb is to reduce projection effects in our measurements. JavaScript movies combining LASCO images with the GOES X-ray light curves (LASCO C2 and LASCO C3 difference: c2rdif_gxray.html and c3rdif_gxray.html) are used to confirm the source locations and obtain the height and width of the propagating CMEs. We used EUV Images and movies from the Sun Earth Connection Coronal and Heliospheric Investigation (SECCHI, Howard et al. 2008) on board the Solar Terrestrial Relations Observatory (STEREO) to detect the eruption locations and confirm the CME-flare connections in SC 24. The observations for SC 23 cover the period from May 1996 to the end of November 2008 (which corresponds to the cycle length of 151 months) whereas SC 24 observations are from the start of December 2008 to the end of November 2019 (cycle length of 132 months). There were an identical number (41) of full HCMEs (CME width, W= $360^0$) in both SC 23 and SC 24. The number of normal CMEs (W<120°) in SC 23 and SC 24 were 445 and 265, respectively. The number of partial HCMEs (120° ≤W< 360°) were similar (114 in SC 23 and 100 in SC 24). The CME speeds extended from 101 km/s to 3163 km/s with a mean value of 694 km/s (SC 23) and from 95 km/s to 2657 km/s with a mean 569 km/s (SC 24). There was no speed measured for the 2006 December 6 HCME due to insufficient data. The soft X-ray flare sizes ranged from a minimum of C3 (selection criterion) to a maximum of X28 (SC 23) and X8.2 (SC 24). The number of limb CMEs associated with the flare sizes of C, M, and X flares were: SC 23 [C3-C9.9: 307, M1-M9.9: 262, X1-X28: 31], and SC 24 [C3-C9.9: 235, M1-M9.2: 148, X1-X8.2: 23]. We measure CME widths as a function of heliocentric distance and compare the width evolutions for a total of 1006 limb CMEs of which 600 are from SC 23 and 406 from SC 24.

## 3. Methodology

We make use of data and measurement tools available at the catalog web site to measure the height and width of the CMEs (https:/cdaw.gsfc.nasa.gov).

### 3.1. Height and angular width measurements

The height of the CME is measured by tracking its leading edge in successive coronagraph images. It is obtained by clicking at the measurement position angle (MPA) of the leading edge. The angular width is measured by clicking subsequently on the edges of the CME at points where the CME intersects the occulting disk. The difference between the two position angles is the width of the CME. Measurement is continued until the CME leading edge is no longer visible (in both C2 and C3 FOV). We identified the faint images of the edges of the CME by playing the JavaScript movies available at the catalog web site.

Figure 1 shows an example event for measuring the height and angular width of a CME that occurred on 2014 May 6 starting at 17:36:05 UT. The limb CME associated with the NOAA active region (AR) 12051 is a partial halo with a speed 815km/s, soft X-ray flare size C4.7, and heliographic location $S09^0$ $W89^0$. The width increased up until a heliocentric distance of 13.8 Rs, stabilized at about 14.25 Rs and then remained constant all the way to 22 Rs.

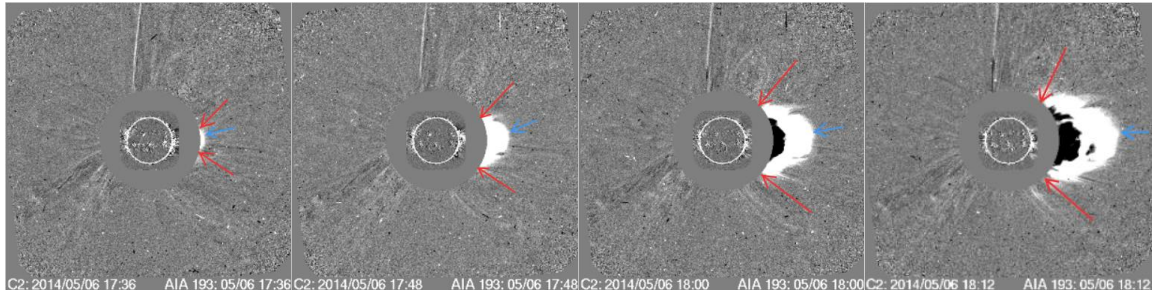

```
#Date            Time            H(Rs)  PA(deg)  Xpix  Ypix    X"        Y"
2014-05-06T17:36:05.481          2.51   275.49   356   277   2401.42    230.86
2014-05-06T17:36:05.481          2.29   283.57   345   289   2139.62    516.46
2014-05-06T17:36:05.481          2.31   257.41   346   247   2163.42   -483.14
2014-05-06T17:48:06.473          3.22   276.51   384   282   3067.82    349.86
2014-05-06T17:48:06.473          2.22   298.42   334   310   1877.82   1016.26
2014-05-06T17:48:06.473          2.21   232.55   326   213   1687.42  -1292.34
2014-05-06T18:00:05.464          4.09   276.16   419   285   3900.82    421.26
2014-05-06T18:00:05.464          2.26   311.80   323   328   1616.02   1444.66
2014-05-06T18:00:05.464          2.27   222.61   317   200   1473.22  -1601.74
2014-05-06T18:12:05.456          5.00   277.89   455   295   4757.62    659.26
2014-05-06T18:12:05.456          2.25   330.29   300   346   1068.62   1873.06
2014-05-06T18:12:05.456          2.21   214.78   306   194   1211.42  -1744.54
```

**Figure 1.** How to measure the width of a CME. The four consecutive frames of the CME llustrate the evolution of the width of the 2014 May 6 17:36:05 UT CME. The red arrows indicate the upper and lower edges of the CME intersecting with the occulting disk. The blue arrow corresponds to the leading edge of the CME. The data output from the measurement tool are given below the images.

### 3.2. Transition heights

We introduce a new parameter which we call the transition height ($hc$) of a CME. We define the transition height as the critical heliocentric distance beyond which the CME width stabilizes to a quasi-constant value. The height immediately after the transition height is the point at which the width maintains a seemingly stable value until the CME leaves the coronagraph FOV [we call this as the height of constant width ($hw$)]. In other words, on the width vs. height graph, the height asymptotes to a constant value.

Figure 2 shows the width vs. height plots for two example events illustrating the way to identify the $hc$ of a CME. The SC 23 limb CME on 2001 Dec 27 at 17:30:05 UT is a normal CME from NOAA AR 9748 (S10$^0$W66$^0$) with a speed: 843 km/s, and a soft X-ray flare size: M2.3. The CME width increased up to $hc$ = 3.5 Rs beyond which it stabilized to a constant value. The SC 24 limb CME which occurred on 2013 Dec 31 at 10:36:05 UT is a partial HCME from NOAA AR 11944 (S09$^0$E101$^0$), speed: 1101 km/s, and a soft X-ray flare size: C8.8. The CME traveled further away from the Sun to acquire its peak width which reached a constant value beyond the heliocentric distance $hc$ = 14.5 Rs.

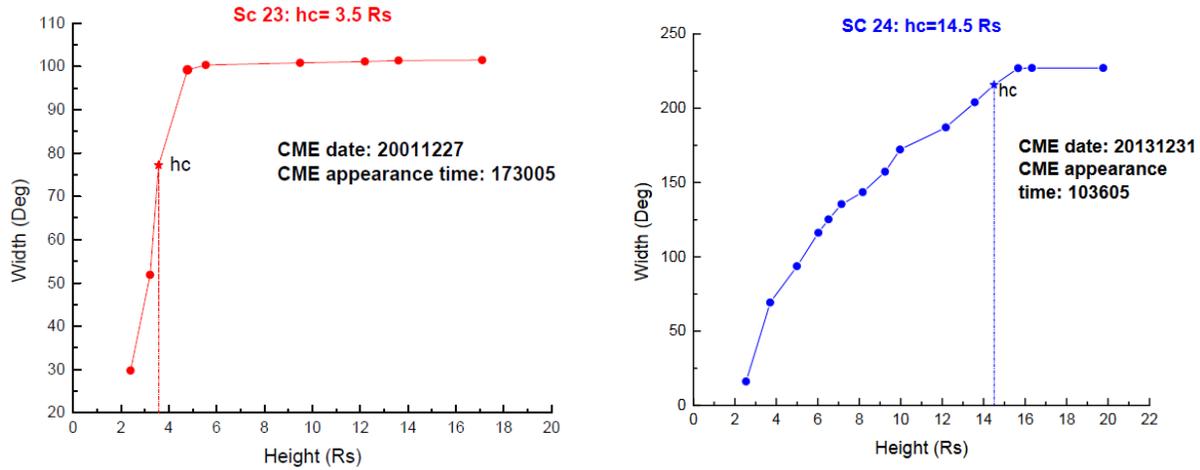

**Figure 2.** The transition height ($hc$) of a SC-23 CME (left) and an SC 24 CME (right) determined from the width vs. height plots. The CME dates, appearance times, and $hc$s are stated in each plot.

## 4. Analysis and Results

We analyze the distributions and variations of $hc$s in SCs 23 and 24. We make use of $hc$ histograms, scatter plots between $hc$ (Rs) and CME date (UT), and Kolmogorov-Smirnov (KS) tests. We compare the width evolution between the two whole cycles and their corresponding phases based on the $hc$ values and their statistical distributions.

### 4.1. Comparison of $hc$s between the two whole solar cycles: SC 23 and SC 24

Figure 3 shows the $hc$ histograms compared between the two cycles. The mean values of $hc$s in SC 23 and SC 24 are 4.7 Rs and 7.6 Rs respectively. The standard error of the mean (SEM) is 0.08 in SC 23 and 0.15 in SC 24. The $hc$ values range from a minimum of 2.21 Rs to a maximum of 21 Rs in SC 23 and from 2.8 Rs to 23 Rs in SC 24 (see Table 1 and 2). The distribution shows that, 61% of the limb CMEs account for $hc < 5$ Rs whereas only 7% make up $hc \geq 7$ Rs in SC 23. In contrast, 55% of the limb CMEs account for $hc \geq 7$ Rs while only 20% for $hc < 5$ Rs in SC 24. The mean, median, standard deviation, and most probable values of $hc$ in SC 24 are higher than the corresponding values in SC 23.

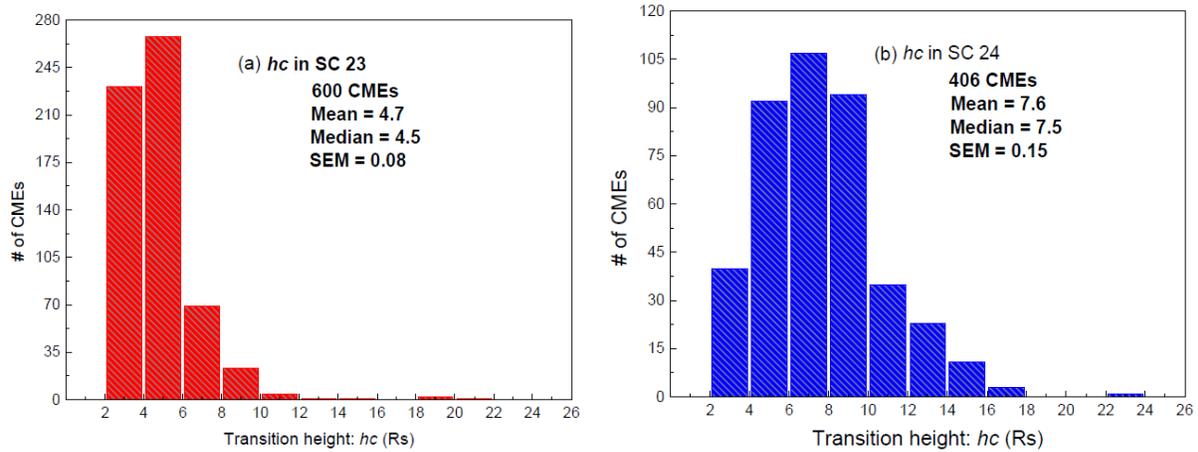

**Figure 3.** Comparison of the transition heights ($hc$s) between SC 23 and SC 24. The red (SC 23) and blue (SC 24) histograms denote the number of limb CMEs with corresponding transition heights. The number of CMEs, the corresponding mean, median, and the standard error of the mean (SEM) values are specified in each plot. The average value of $hc$ in SC 24 is larger than in SC 23. The data used to create figure 3(a) and 3(b) are available in the machine readable formats of Table 1 and Table 2.

Table 1 and 2 show the contents of SC 23 and SC 24 limb CMEs: CME appearance date and time (columns 1-6); CME width, speed, and position angle (columns 7-9); $hc$ and $hw$ (columns 10-11). Table 1 lists the flare on set date and time, and the flare peak time in columns 13-19 while Table 2 lists same in columns 12-18. The soft X-ray flare size, source location, and active region number are given in columns 20-22 (Table 1) and columns 19-21 (Table 2). The apparent $hw$ in column 11 (Table 1) indicates the hw of some CMEs whose actual $hw$ is difficult to measure; the actual $hw$ values were estimated from interpolation/extrapolation as indicated in column 12 (Table 1).

**Table 1**

Contents of solar cycle 23 limb CMEs

| Num | Units | Label | Explanations |
| --- | --- | --- | --- |
| 1 | yr | CME.Y | UT Year of CME appearance |
| 2 | m | CME.M | UT Month of CME appearance |
| 3 | d | CME.D | UT Day of CME appearance |
| 4 | h | CME.h | UT Hour of CME appearance |
| 5 | min | CME.m | UT Minute of CME appearance |
| 6 | s | CME.s | UT Second of CME appearance |
| 7 | deg | Width | CME width |
| 8 | km/s | Speed | CME speed (speed immeasurable for the 2006/12/06 event) |
| 9 | deg | PA | CME position angle |
| 10 | solRad | hc | CME transition height |
| 11 | solRad | hw | CME height of constant width (*Apparent hw) |
| 12 | solRad | hwint. | CME height of constant width (**Interpolated) |
| 13 | yr | Flare.Y | UT Year of flare onset |
| 14 | m | Flare.M | UT Month of flare onset |
| 15 | d | Flare.D | UT Day of flare onset |
| 16 | h | FlareS.h | UT Hour of flare onset |
| 17 | min | FlareS.m | UT Minute of flare onset |
| 18 | h | FlareP.h | UT Hour of flare peak |
| 19 | min | FlareP.m | UT Minute of flare peak |
| 20 | -- | Size | Soft X-ray flare size |
| 21 | deg | Location | CME Source location |
| 22 | -- | AR | Active Region |

This table is available in its entirety in the machine readable format. Only a portion of the table is shown here to demonstrate its form and content.

**Table 2**

Contents of solar cycle 24 limb CMEs

| Num | Units | Label | Explanations |
|---|---|---|---|
| 1 | yr | CME.Y | UT Year of CME appearance |
| 2 | m | CME.M | UT Month of CME appearance |
| 3 | d | CME.D | UT Day of CME appearance |
| 4 | h | CME.h | UT Hour of CME appearance |
| 5 | min | CME.m | UT Minute of CME appearance |
| 6 | s | CME.s | UT Second of CME appearance |
| 7 | deg | Width | CME width |
| 8 | km/s | Speed | CME speed |
| 9 | deg | PA | CME position angle |
| 10 | solRad | hc | CME transition height |
| 11 | solRad | hw | CME height of constant width |
| 12 | yr | Flare.Y | UT Year of flare onset |
| 13 | m | Flare.M | UT Month of flare onset |
| 14 | d | Flare.D | UT Day of flare onset |
| 15 | h | FlareS.h | UT Hour of flare onset |
| 16 | min | FlareS.m | UT Minute of flare onset |
| 17 | h | FlareP.h | UT Hour of flare peak |
| 18 | min | FlareP.m | UT Minute of flare peak |
| 19 | -- | Size | Soft X-ray flare size |
| 20 | deg | Location | CME Source location |
| 21 | -- | AR | Active Region |

This table is available in its entirety in the machine readable format. Only a portion of the table is shown here to demonstrate its form and content

The scatter plots of $hc$s as a function of CME dates in Fig. 4 show that the values of $hc$ in SC 24 are significantly higher than in SC 23. There are 600 and 406 limb CMEs during the whole cycle periods in SC 23 (1996 May 1-2008 November 30) and SC 24 (2008 December 1-2019 November 30), respectively.

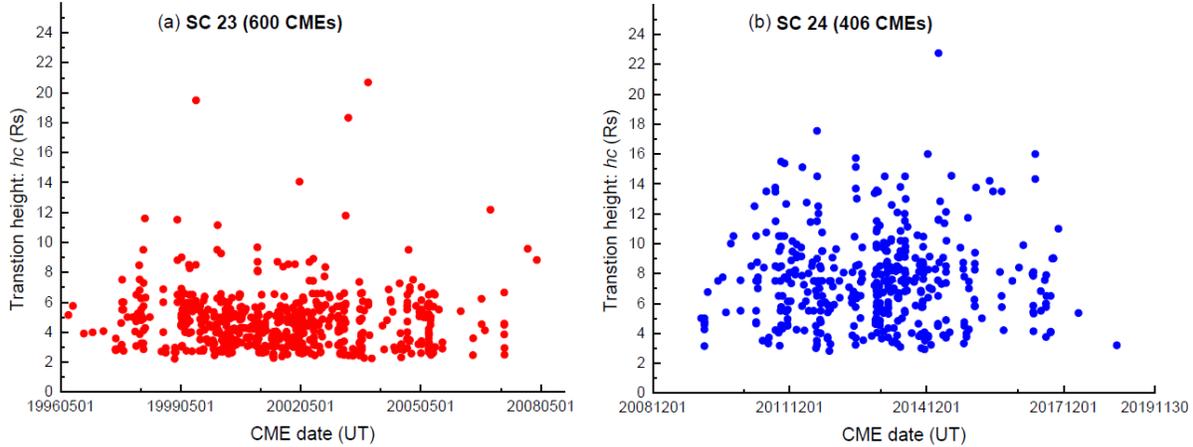

**Figure 4.** Transition heights as a function of CME dates showing the time variations of 600 limb CMEs in SC 23 (left) and 406 limb CMEs in SC 24 (right). Figure 4(a) and 4 (b) can be reproduced using the data available in the machine readable formats of Table 1 and Table 2.

We performed a two-sample Kolmogorov-Smirnov (KS) test to confirm the statistical significance of the difference in hcs between the two data sets in SC 23 and SC 24 (https://www.sci.utah.edu/~arpaiva/classes/UT_ece3530/hypothesis_testing.pdf).
Table 3 shows a summary of the KS test results for the *hc*s between the two whole solar cycles and their corresponding phases. KS test uses the statistic D, which is the maximum absolute difference between the observed cumulative distribution functions of the two data sets compared. When the KS test statistic D exceeds a critical value (Dc) the null hypothesis (H0) that the two samples belong to the same distribution is rejected. For sample sizes >12, $D_C$ is given by the formula: $D_C = c(\alpha)\sqrt{(n+m)/nm}$ where n and m are the sizes of the two samples compared. Here c(α) is a constant for a given level of significance α. For α = 0.05, c(α) = 1.36 which we use in this study. Comparing whole cycles, n = 600 (SC 23) and m = 406 (SC 24), so we get $D_C$ = 0.087. While comparing the *hc*s of SC 23 and 24, we get a D statistic (0. 518) that far exceeds Dc = 0.087. Therefore, the H0 is rejected. The KS test implies that the difference between the *hc* distributions of the two data sets is highly statistically significant (p = 0). The mean, median, standard deviation, and maximum values of *hc* are significantly higher in SC 24 compared to those in SC 23. Similar analysis is performed for other sample sets and the results are given in Table 3**.**

**Table 3**

Kolmogorov-Smirnov (KS) test result summary

| Data set | | | P-value | D | $D_C$ | Result |
|---|---|---|---|---|---|---|
| **CME parameter** | **Comparisons** | **Phase** | | | | |
| Transition heght ($hc$) | SC 23 and SC 24 | Whole cycle | 0 | 0.518 | 0.087 | Statistically significant. |
| | Intercycle comparison. (SC 23 and SC 24) | Rising | 0.158 | 0.33 | 0.397 | Not statistically significant |
| | | Maximum | 0 | 0.561 | 0.115 | Statistically significant. |
| | | Decline | 0 | 0.519 | 0.185 | |
| | Intracycle (SC 23 only) | Rise and maximum | 0.292 | 0.145 | 0.201 | Not statistically significant. |
| | | Maximum and decline | 0.252 | 0.089 | 0.119 | |
| | | Rise and decline | 0.486 | 0.128 | 0.208 | |
| | Intracycle (SC 24 only) | Rise and maximum | 0.235 | 0.275 | 0.361 | Not statistically significant. |
| | | Maximum and decline | 0.796 | 0.068 | 0.142 | |
| | | Rise and decline | 0.277 | 0.269 | 0.368 | |
| X-ray flare size | X-ray class | Whole cycle | | | | |
| | C | | 0.282 | 0.086 | 0.118 | Not statistically significant |
| | M | | 0.998 | 0.04 | 0.14 | |
| | X | | 0.349 | 0.257 | 0.374 | |
| CME speed | SC 23 and SC 24 | Whole cycle | 0.0007 | 0.128 | 0.0087 | Statistically significant. |

4.2. Comparison of *hc*s between the solar phases of SC 23 and SC 24

We conducted the intercycle and intracycle comparisons of *hc*s in the corresponding phases (rise, maximum, and declining) of the two cycles. Figure 5 shows the variations of the transition heights over time and a comparison of *hc*s distributed over the corresponding phases (rise, maximum, and declining) of the solar cycles. The *hc* distributions in the rising phases of the two cycles are almost similar. There appears an appreciable difference in the maximum and declining phases between the two cycles with the highest divergence pronounced in the maximum phase.

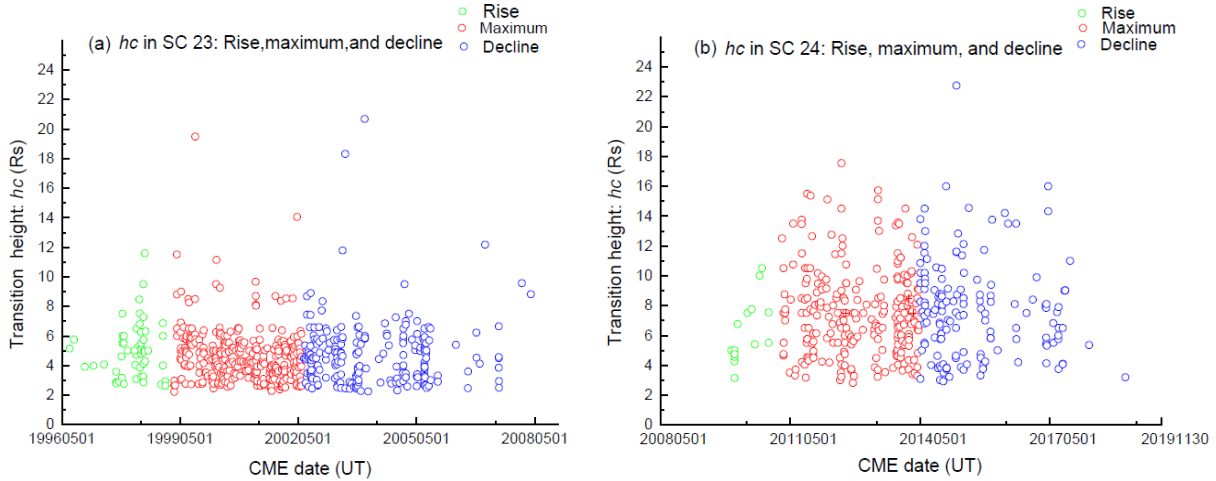

**Figure 5.** Variations of *hc*s over time for the corresponding phases of SC 23 (left) and SC 24 (right). The green, red, and blue circles refer to the rising, maximum, and declining phases in both cycles. The data used to create figure 5(a) and 5(b) are available in the machine readable formats of Table 1 and Table 2.

There are very few limb CMEs in the rising phase of the two cycles, but slightly more in SC 23. The declining phase accounts for 36% of the limb CMEs in each cycle. The maximum phase constitute 60% of the 406 limb CMEs in SC 24 compared to the 55% of the 600 limb CMEs in SC 23. The *hc* histograms in Fig.6 show comparisons with in the phases of each solar cycle and between the phases of the cycles. In the maximum phase, about 88% of the limb CMEs in SC 23 and only 32% of those in SC 24 account for $hc < 6$ Rs. But then again, in the same phase, 55% of the limb CMEs in SC 24 and only 6% of those in SC 23 make up $hc \geq 7$Rs. There are no significant differences in the average values of *hc*s between the rising phases of the two cycles.

In the maximum and declining phases, the mean and median values of *hc*s in SC 24 are significantly higher than in SC 23.

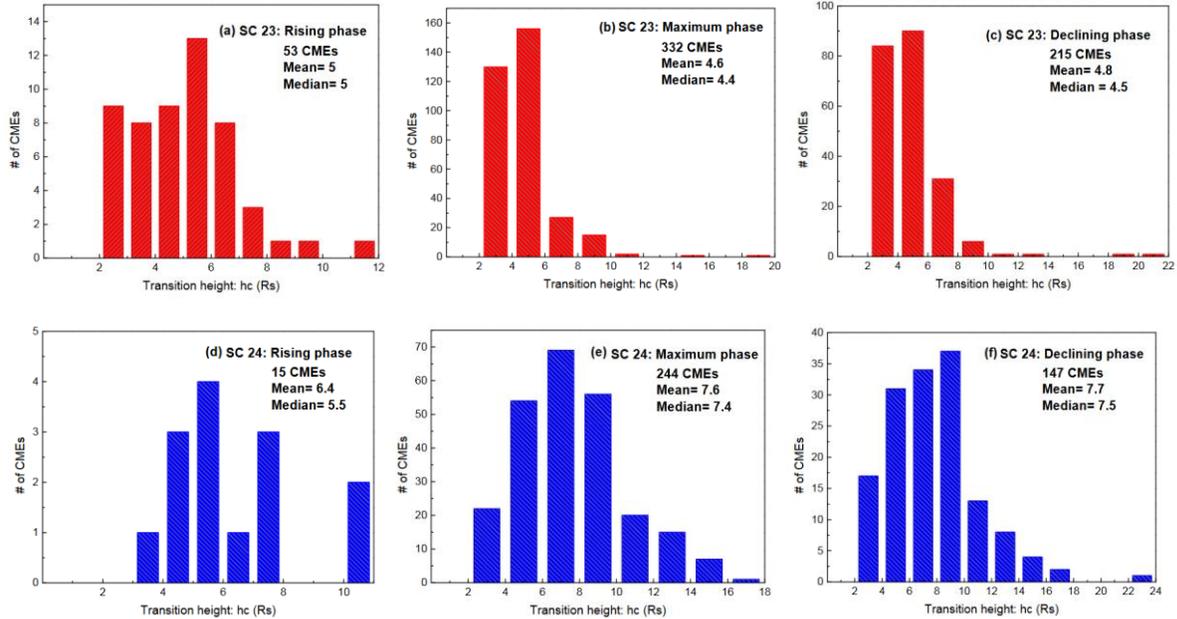

**Figure 6:** Comparison of the transition heights between the corresponding phases of SC 23 and SC 24. The red (top row: a, b, and c) and blue (bottom row: d, e, and f) histograms correspond to the number of limb CMEs Vs. transition heights for the rise, maximum, and declining phases of SC 23 and SC 24, respectively. The number of limb CMEs in the rising phase (left-hand panels), maximum phase (middle panels), and declining phase (right-hand panels) along with the corresponding mean and median values are plotted on the panels. Figure 6 (a, b, c) and (d, e, f) can be reproduced using the data available in the machine readable formats of Table 1 and 2.

We also performed the KS test for the phase-to-phase comparison of *hc*s between the two solar cycles (see Table 3). The KS test in the rising phase ($D = 0.33 < D_C = 0.397$ and $p = 0.158$) implies that the *hc*s between the two cycles are not significantly different. The maximum ($D = 0.561 > D_C = 0.115$ and $p = 0$) and declining phases ($D = 0.519 > D_C = 0.185$ and $p = 0$) show that the difference in *hc*s between the two cycles is statistically significant. The *hc*s in the maximum and declining phases of SC 24 are significantly higher than the corresponding phases in SC 23. We also conducted the KS test within the phases of each solar cycle. The KS tests in the phases of SC 23 [maximum and decline ($D = 0.089 < D_C = 0.119$ and $p = 0.252$), maximum and rise (D = 0.145 < $D_C$ = 0.201 and p = 0.292), and decline and rise (D = 0.128 < $D_C$ = 0.208 and p = 0.486)] imply that the *hc* differences are not statistically significant. In the phases of SC 24, the KS tests [maximum and decline (D = 0.068 < $D_C$ = 0.142 and p = 0.796), maximum and rise (D = 0.275 < $D_C$ = 0.361 and p = 0.235), and decline and rise (D = 0.269 < $D_C$ = 0.368 and p = 0.277)] also signify that the *hc* values are not significantly different. There is no significant difference in the *hc* values within the phases of each solar cycle.

### 4.3. Flare size distributions

We analyze the flare size distributions between the two cycles in order to test whether the source properties are responsible for the *hc* differences. We selected limb CMEs with flare sizes ≥ C3 so as to avoid the uncertainties in the CME identification for weak flares. Figure 7 shows a comparison of the X-ray flare sizes related with the corresponding sets of limb CMEs. The average flare sizes in the two cycles are similar: M3.8 (SC 23), M2.7 (SC 24). The C flares account for the highest percentage in each cycle. The X flares have similar fractions in both cycles (SC 23: 5%, SC 24: 6%). The percentages of C and M flares between the two cycles are also similar: SC 23 (C: 51%, M: 44%) and SC 24 (C: 58%, M: 37%). The mean and median values of the C and M flares in the two cycles are similar. We performed a two sample KS test for each X-ray class (see Table 3). The KS test results (D < $D_C$ and p > α) imply that there is no statistically significant difference in the soft X-ray flare sizes between SC 23 and SC 24.

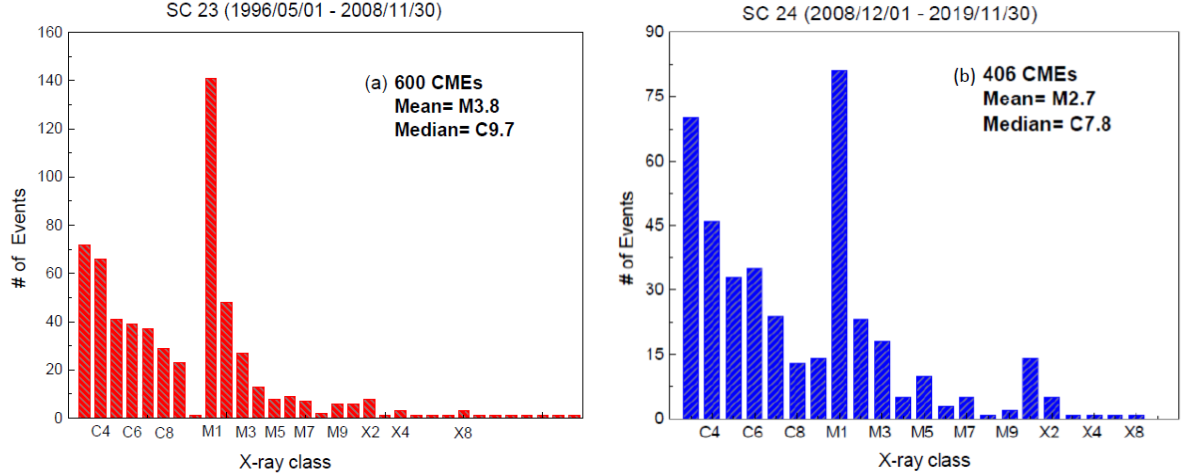

**Figure 7.** Size distributions of GOES soft X-ray flares associated with CMEs in SC 23 (left) and SC 24 (right). The number of limb CMEs, the mean and median values of the X-ray flare sizes of each cycle are plotted on the panels. Figure 7(a) and 7(b) can be reproduced using the data available in the machine readable formats of Table 1 and Table 2.

### 4.4. Limb CME kinematics
#### 4.4.1 Speed comparisons

The selection of CMEs close to the solar limb (within $30^0$ of the limb) diminishes projection effects in our speed measurement, so no correction for the sky plane speeds was made. Thus, we consider the sky plane speeds as the true speeds of the limb CMEs. Figure 8 shows a comparison of the limb CME speeds between the two cycles. The average speed in SC 23 (694 km/s) is 22% higher than that in SC 24 (569 km/s). The speed distributions in each cycle are lognormal with median values 554 km/s in SC 23 and 450 km/s in SC 24. The fraction of the limb CMEs which account for speeds $\leq$ 500 km/s in each cycle are nearly identical (SC 23: 56%, SC 24: 55%). Only 3% of the CMEs in SC 24 and 7% of those in SC 23 account for speeds $\geq$ 1500 km/s. A two sample KS test for the speeds (see Table 3) generates a D statistic = 0.128 with a p value = 0.0007 ($D > D_C$ and $p < \alpha$) which implies that the speeds between the two cycles differ significantly. These results are consistent with the speed difference observed in the general population of CMEs and limb halo CMEs (Gopalswamy et al. 2020a, b).

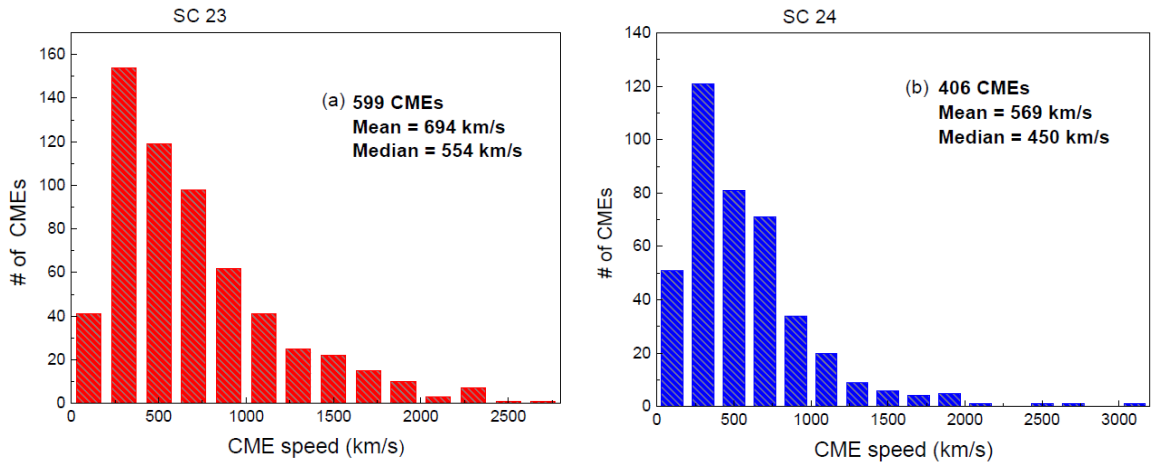

**Figure 8**. Comparison of speeds of limb CMEs in SCs 23 (left) and 24 (right). The number of limb CMEs and the corresponding mean and median CME speeds are stated in each plot. Figure 8(a) and 8 (b) can be reproduced using the data available in the machine readable formats of Table 1 and Table 2.

### 4.4.2. Speed Vs. transition heights

The scatter plots between the limb CME speeds and $hc$s for a total of 1002 limb CMEs (SC 23: 596, SC 24: 406) are shown in Figure 9. In SC 23, 4 CMEs out of the 600 were excluded from the analysis. The 2006 December 6 CME had no measured speed due to inadequate data. CMEs dated on 1999/09/19 17:18:05 UT, 2003/07/10 16:33:07 UT, and 2004/01/06 08:53:06 UT which appeared as outliers in the plots were dropped because their early evolution is missed (see Section 5 for details). The regression lines, correlation coefficients, slopes, and intercepts are shown on the plots. The Pearson critical correlation coefficients ($r_c$) for the sample sizes of 406 and 596 are 0.098 and 0.088, respectively. The correlation coefficient in SC 24 (r = 0.67> $r_c$ = 0.098) indicates a stronger, significant and positive relationship between the speeds and their corresponding $hc$s. In SC 23, although the correlation is significant (r = 0.47> $r_c$ = 0.088), it is weaker. The slope of the regression line in SC 24 (0.0048) is significantly higher than in SC 23 (0.0017). The SC 24 regression line is steeper by 182% which signifies that, for a given CME speed, the transition height in SC 24 is higher than in SC 23. The data points are clustered in the lower left (SC 23) and upper left (SC 24) of the plot with most SC 23 $hc$s in the 4-5 Rs range and in SC 24 in the 8-12 Rs range. The regression line in SC 24 gets steeper with the increase in CME speeds. The difference in the *hc*s between the two cycles is more appreciable at higher speeds (>1000 km/s).

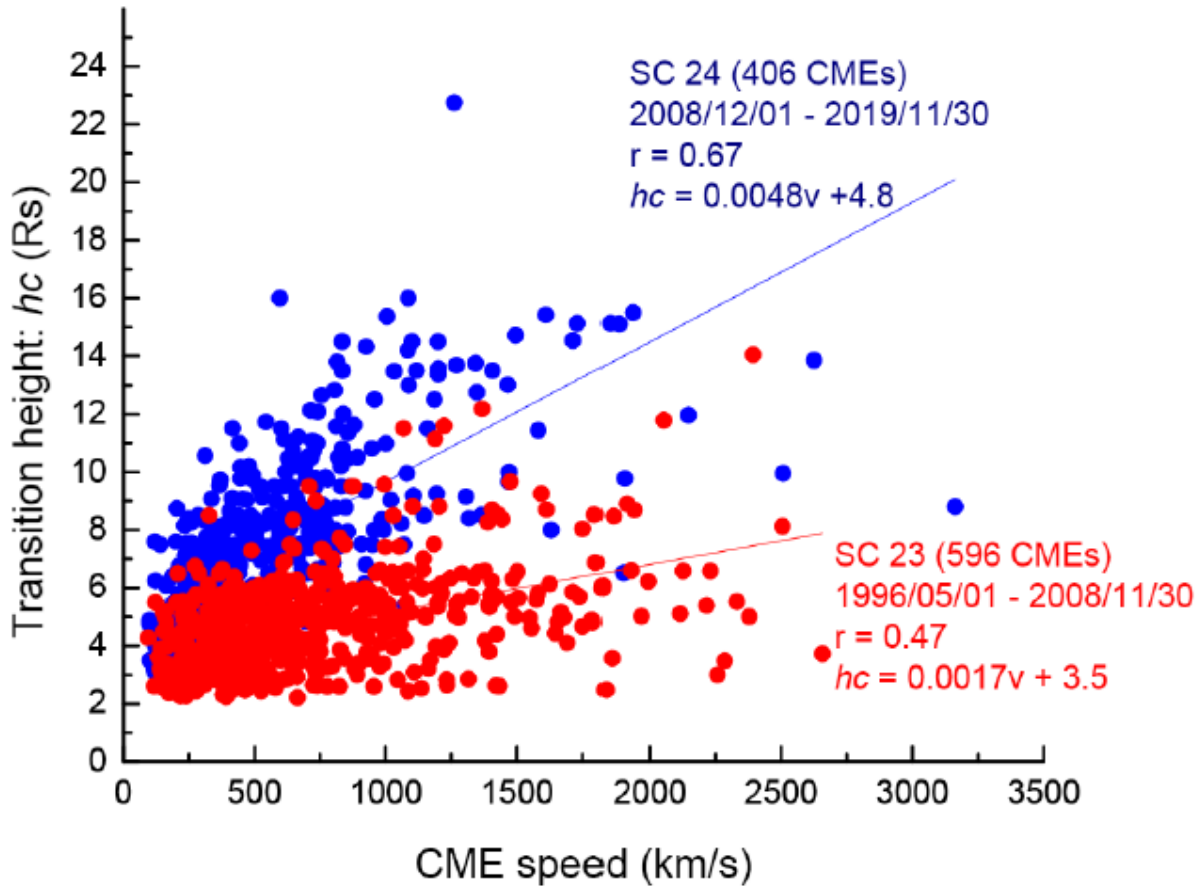

**Figure 9**. Scatter plots between the CME speeds and transition heights of 1002 limb CMEs in SC 23 (red) and SC 24 (blue). The red and blue solid lines signify linear fit to the data points. The slope of the regression lines and the corresponding equations along with the correlation coefficients are shown in the plots. Note the distinct regions occupied by the SC 23 and SC 24 data points on the plots. The data used to create this figure are available in the machine readable formats of Table 1 and Table 2.

## 5. Discussion

We have made comparisons of the width evolution of the limb CMEs based on the transition heights. The CME maintains a seemingly stable width value at the height of constant width (*hw*) until it leaves the coronagraph FOV. The *hw*s of most of the CMEs are slightly higher than their corresponding *hc*s. However, some CMEs have significantly higher *hw*s. Noticeable differences between the two heights may occur when the CME leading edge moves beyond the LASCO/C2 FOV at the time of constant width which makes it difficult to measure the actual value of *hw*. This may happen due to low cadence in SC 23. In view of this, one might question whether our conclusions are valid if the comparisons are based on *hw*s. We have identified 34 CMEs of such type in SC 23. We addressed the solution to this problem from two perspectives: I. We dropped the 34 CMEs. II. We maintained the CMEs and implemented the technique of Interpolation/extrapolation to estimate the actual *hw* values. Furthermore, we excluded four CME events in SC 23. There was no speed measured for the CME on 2006/12/06 20:12:05 UT. The CME on 1999/09/19 17:18:05 UT had a LASCO/C2 data gap (1999/09/19 17:06-1999/09/19 18:33); measurements were only in C3 FOV and early width evolution was not determined. On 2003 July 10 16:33:07 UT, the LASCO/C2 data gap just ended but the CME front was beyond C2 FOV when observations resumed. The 2004 January 06 08:53:06 UT CME also had a data gap (2004/01/05 21:40-2004/01/06 09:20) and only 2 height-time points in C3 FOV. There were no observations for the initial width evolutions and the heights are overestimated. Thus, these 3 events which appeared as outliers in both plots (Figures 10 and 11) were removed. Consequently, the number of limb CME events in SC 23 for the *hw* analysis from the view points of I and II are 562 and 596, respectively.

A comparison of the *hw*s shows that the mean, median, and SEM values of *hw* in SC 24 (8.42 Rs, 8.04 Rs, and 0.17, respectively) are significantly higher than the corresponding values in SC 23. When we drop the 34 CMEs: the mean, median, and SEM values of *hw* in SC 23 are 5.63, 5.35, and 0.1, respectively. When we implement interpolation/extrapolation: the mean, median, and SEM values are 5.89, 5.47, and 0.1, respectively. The KS tests for the *hw*s between the two cycles [D = 0.46 > $D_C$ = 0.089, p = 0 (when dropped) and D = 0.414 > $D_C$ = 0.087, p = 0 (during interpolation/extrapolation)] show that the *hw* differences are statistically significant. The scatter plots between the CME speeds and the *hw*s are shown in Figure 10 (the 34 CMEs were dropped) and Figure 11 (*hw*s estimated from interpolation/extrapolation). Figure 10 shows that the slope of the regression line in SC 24 (0.0059) is significantly higher than in SC 23 (0.0022). The SC 24 regression line is steeper by 168% implying that, for a given CME speed, the *hw* in SC 24 is higher than in SC 23. Figure 11 shows that the slope of the regression line in SC 24 (0.0059) is 111% higher than in SC 23 (0.0028). In either way, we obtained similar results in terms of correlation coefficients and statistical parameters.

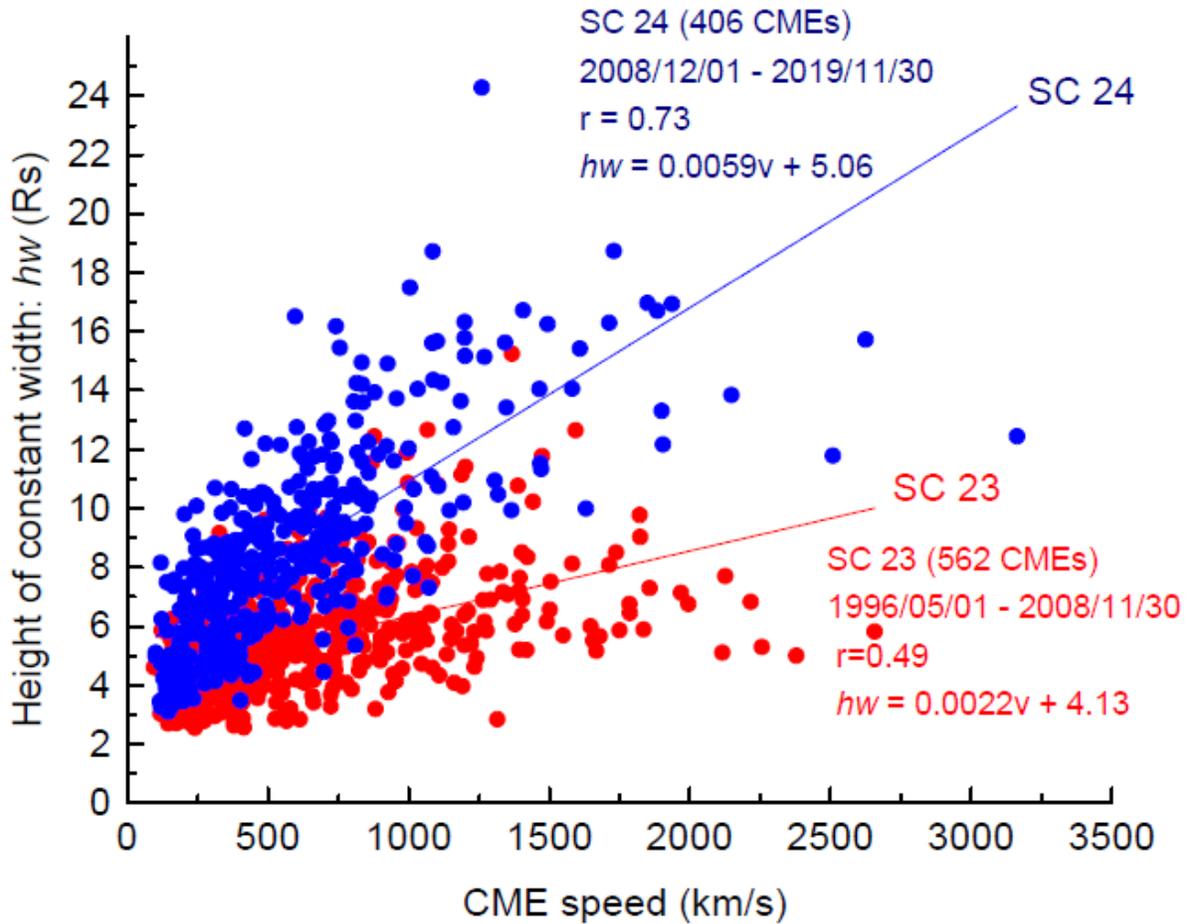

**Figure 10**. Scatter plots between the CME speeds and the heights of constant width of limb CMEs in SC 23 (red) and SC 24 (blue). There are only 562 CMEs in SC 23. The 34 CMEs whose actual $hw$s are difficult to measure were dropped. The slope of the regression line in SC 24 is 168% higher than in SC 23. The data used to create this figure are available in the machine readable formats of Table 1 and Table 2.

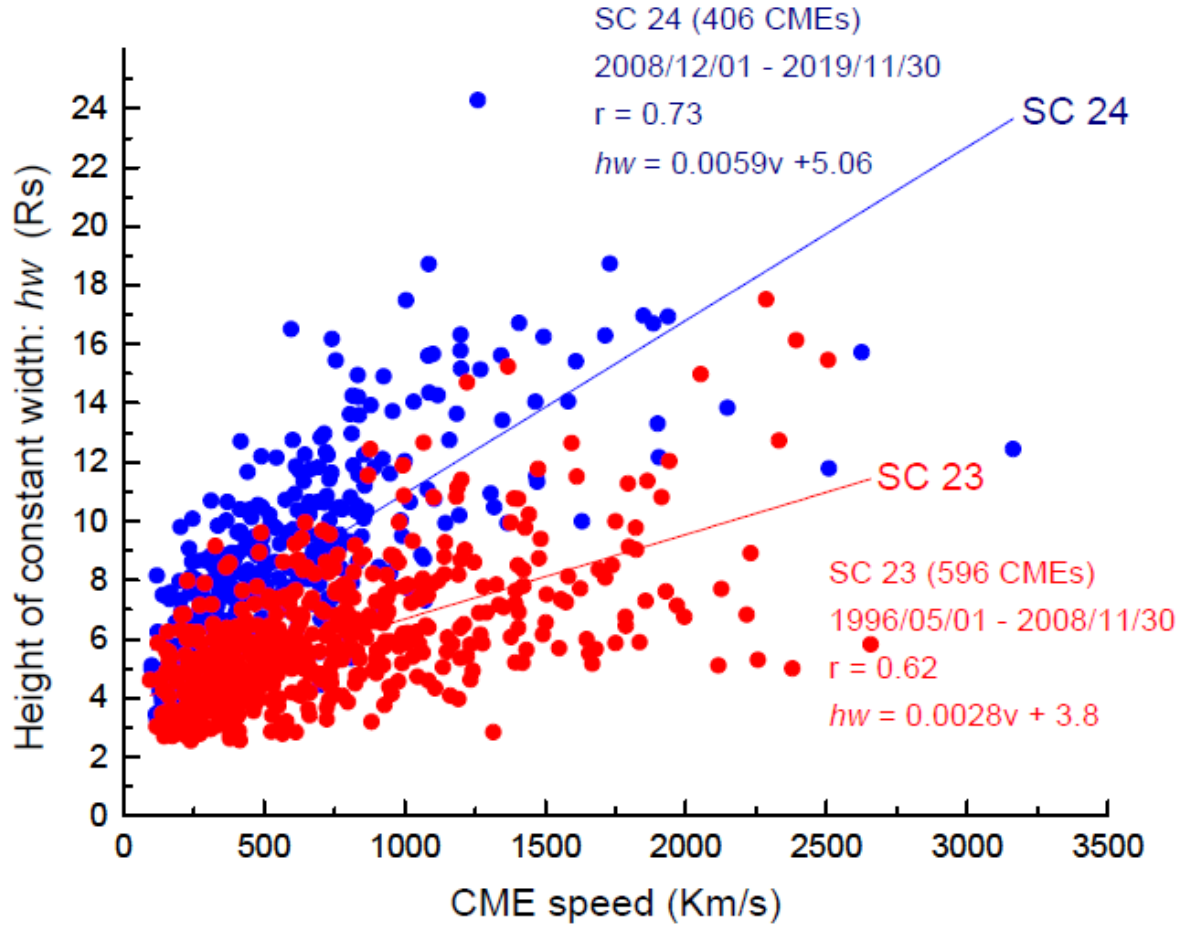

**Figure 11**. Scatter plots between the CME speeds and the heights of constant width. There are 596 CMEs in SC 23. Interpolation/extrapolation is used to estimate the actual *hw* values of the 34 CMEs. The slope of the regression line in SC 24 is 111% higher than in SC 23. The data used to create this figure are available in the machine readable formats of Table 1 and Table 2.

## 6. Summary and Conclusions

The culmination of solar cycle 24 by the end of 2019 has created the opportunity to compare the differing properties of CMEs between two complete solar cycles (SC 23 and SC 24). We analyzed significantly large numbers of limb CMEs in SC 23 (1996 May 1-2008 November 30: 600 CMEs) and in SC 24 (2008 December 1-2019 November 30: 406 CMEs). We measured CME width as a function of heliocentric distance and determined the distances where the CMEs reach constant width in each solar cycle. We introduced a new parameter named as the transition height ($hc$) of a CME and defined it as the critical heliocentric distance beyond which the CME width stabilizes to a quasi-constant value. We have compared the width evolution of limb CMEs in the two whole cycles and their corresponding phases (rise, maximum, and declining) based on this new parameter. A comparison of $hc$s using statistical distributions and KS tests confirmed that the difference between the $hc$ distributions of the two data sets is highly statistically significant. We have also compared the $hc$s within the phases of each cycle and between the phases of the two cycles. Our comparison confirmed that there is no significant difference in the $hc$ values within the phases of a given solar cycle. The $hc$s between the phases of the two cycles are significantly different. The flare size distributions between the two cycles do not differ significantly which proves that it is unlikely for the source properties to have caused the $hc$ differences. The CME speeds between the two cycles differ significantly. For a given CME speed, $hc$ in SC 24 is significantly higher than in SC 23. We have proved that the $hc$s in SC 24 are significantly higher than in SC 23 which implies that SC 24 CME flux ropes attain their peak width at a larger heliocentric distance than the SC-23 flux ropes did. Thus, our analysis confirms the prediction by Gopalswamy et al. (2015a) based on the apparent contradiction between CME widths at the Sun and flux rope sizes at 1 au. Furthermore, the confirmation of the prediction points to the importance of the total pressure difference between the flux rope and ambient medium.

The specific conclusions of this work are summarized as follows:

1. We introduced a new parameter: the transition height ($hc$) of a CME, defined as the critical heliocentric distance beyond which the CME width stabilizes to a quasi-constant value. The heliocentric distance point immediately after the transition height is the point at which the width maintains a seemingly stable value.
2. The average $hc$ in SC 24 (7.6 Rs) is significantly higher than that in SC 23 (4.7 Rs).
3. A comparison of $hc$s between the two whole cycles proves that the difference between the $hc$ distributions of the two data sets is highly statistically significant. The larger values of $hc$s in SC 24 signify that cycle 24 CMEs attain their peak width at larger distances from the Sun as compared to cycle 23 CMEs. Thus, the heliocentric distance for the pressure balance between the flux rope and the ambient medium is higher in SC 24 than in SC 23.

4. The intercycle comparisons of the transition heights imply that the difference in *hc*s between the two cycles is statistically significant. The *hc*s in the maximum and declining phases of SC 24 are significantly higher than the corresponding phases in SC 23. The intracycle comparisons show that the *hc* values within the phases of a given cycle do not differ significantly.
5. The average flare size in SC 23 (M3.8) is similar to that in SC 24 (M2.7). The size distributions of the associated flares between the two cycles are also similar with median values of C9.7 (SC 23) and C7.8 (SC 24). There is no statistically significant difference in the soft X-ray flare sizes between SC 23 and SC 24. The implication is that the source properties don't account for the differences in *hc* between the two cycles.
6. The limb CME kinematics between the two cycles differ significantly. The average speed in SC 23 (694 km/s) is 22% higher than that in SC 24 (569 km/s). This is consistent with the results obtained for the general population of CMEs and limb halo CMEs.
7. There is a significant correlation between the limb CME speeds and their corresponding *hc*s. The correlation in SC 24 is stronger with correlation coefficient (r = 0.67) but relatively weaker in SC 23 (r = 0.47). The slope of the regression line in SC 24 is 182% higher than in SC 23. The steeper regression line in SC 24 implies that for a given CME speed, the *hc* in SC 24 is significantly higher than in SC 23. The *hc* differences are progressively noticeable with the rise in CME speeds.

The diminished solar activity in SC 24 demonstrated by a substantial decline in the heliospheric parameters of great significance in the interaction of the heliosphere with interstellar medium has resulted in the weak state of the heliosphere. The anomalous expansion of cycle 24 CMEs caused by the weak state of the heliosphere accounts for the larger heliocentric distance where the pressure balance between CME flux rope and the ambient medium is attained. The enhanced transition height in SC 24 is new observational confirmation of the anomalous expansion. This work which reveals the effect of the heliospheric state on CME width evolution is another manifestation of the diverse interaction between the heliosphere and CMEs in the two cycles.

**Acknowledgment**

We acknowledge NASA's open data policy in using SOHO and STEREO, and NOAA's GOES X-ray data. SOHO is a project of international collaboration between ESA and NASA. STEREO is a mission in NASA's Solar Terrestrial Probes program. This work is supported by NASA's Living With a Star program. Most of this work was carried out when F.D. visited NASA Goddard Space Flight Center.

**ORCID iDs**

Fithanegest Kassa Dagnew https://orcid.org/0000-0003-0817-5983

Nat Gopalswamy https://orcid.org/0000-0001-5894-9954

Sachiko Akiyama https://orcid.org/0000-0002-7281-1166

Seiji Yashiro https://orcid.org/0000-0002-6965-3785